\documentclass[12pt]{article}

\usepackage{scicite}
\usepackage{graphicx} 

\usepackage{times}

\newenvironment{sciabstract}{%
\begin{quote} \bf}
{\end{quote}}

\newcounter{lastnote}
\newenvironment{scilastnote}{%
\setcounter{lastnote}{\value{enumiv}}%
\addtocounter{lastnote}{+1}%
\begin{list}%
{\arabic{lastnote}.}
{\setlength{\leftmargin}{.22in}}
{\setlength{\labelsep}{.5em}}}
{\end{list}}

\title{Molecular imaging with X-ray free electron lasers: dream or reality?}

\author
{Andrea Fratalocchi,$^{1\ast}$ Giancarlo Ruocco$^{1,2}$\\
\\
\normalsize{$^{1}$Department of Physics, Sapienza University,}\\
\normalsize{P.le Aldo Moro 2, 00185, Rome, ITALY}\\
\normalsize{$^{2}$IPCF-CNR, c/o Department of Physics, Sapienza University,}\\
\normalsize{P.le Aldo Moro 2, 00185, Rome, ITALY}\\
\\
\normalsize{$^\ast$E-mail:  andrea.fratalocchi@uniroma1.it.}
}

\date{}

\begin{document} 


\baselineskip24pt


\maketitle


\begin{sciabstract}
X-ray Free Electron Lasers (XFEL) are revolutionary photons sources, whose ultrashort, brilliant pulses are expected to allow single molecule diffraction experiments providing structural information on the atomic length scale. This ultimate goal, however, is currently hampered by several challenging questions basically concerning sample damage, Coulomb explosion and the role of nonlinearity. By employing an original \emph{ab-initio} approach, as well as exceptional resources of parallel computing, we address these issues showing that accurate XFEL-based single molecule imaging will be only possible with ultrashort pulses of half of femtosecond, due to significant radiation damage and the formation of preferred multi-soliton clusters which reshape the overall electronic density of the molecular system at the femtosecond scale.

\end{sciabstract}


\paragraph*{Introduction}
Within the next few years three X-ray Free Electron Lasers (XFEL), namely the Linac Coherent Light Source (LCLS) at Stanford \cite{lcls}, the European XFEL in Hamburg \cite{xfele} and the Japanese XFEL at RIKEN \cite{xfelj}, will start the operations. These facilities will generate the brightest X-ray pulses in the world, with atomic scale wavelength ($>0.05$ nm) and maximum peak powers ($0.1-0.5$ TW) order of magnitudes beyond the current capabilities of synchrotron-based X-ray sources. Throughout history, the discovery of new light sources has given rise to decisive steps forward in science, and the new XFEL facilities hold the promise of making accessible a new fundamental physical domain, thus establishing a new era in science and research. This has led to the flourishing of an incredibly large interest around the XFEL in recent years \cite{science_rev_chap,nat_xfel_j,fel_fusion,sc_chem,sc_rev,rev_xfel_chap,chap_del,xfel_int,db_vis,fl_micr,PhysRevLett.92.074801,krausz:163}. Among the various challenges proposed ---which range from extreme investigations in high-energy physics to ultrafast chemical analysis--- the possibility to perform coherent diffraction imaging at the atomic scale has stirred particular interest \cite{science_rev_chap,sc_chem,rev_xfel_chap,chap_del,db_vis,fl_micr,caght_spin,caghtII}. The difficulties in determining the structure of proteins that cannot be crystallized, in fact, is one of the leading challenges for structural biology today, and the extremely brilliant, ultrashort, bursts of the XFEL have the potential to get around this problem; on the one hand the XFEL pulse duration (femtoseconds), is the natural time scale of elementary atomic interactions and it is expected to open the study of the evolution of fundamental processes, which could otherwise be studied only indirectly; on the other the XFEL could have the power to enable diffraction imaging of single molecules, thus overcoming conventional crystallography that works essentially through linear amplification (by observing discrete Bragg peaks resulting from the interference of regular arrangement of atoms). The initial steps for this new frontier of research have begun at the FLASH facility \cite{flash}, where imaging experiments have successfully captured a time-series of snapshots showing the nonequilibrium, ultrafast transient dynamics of a nanostructured solid: the laser ablation of a silicon film \cite{barty_abla} ---with a spatial resolution of $50$ nm and a temporal resolution of $10$ ps--- and the explosion dynamics of polystyrene spheres with diameter of $140$ nm \cite{chap_del}. However, to reach the ultimate goal of single molecule imaging, things turn out to be more complicated. Aside from the most fundamental questions concerning the phenomenology arising from high intensity photon beam-matter interactions (which are not fully exploited) there is a practical aspect that needs to be taken into account. Atomic scale imaging of particles, in fact, will be lensless, with the far field diffraction pattern scattered by the molecule first recorded on a ÔcameraÕ and then analyzed through a computer algorithm, which reconstructs the molecular image \cite{science_rev_chap,schroer:090801,imag_lensless_met,len_imag:eise,fl_micr,caght_spin,caghtII}. To theoretically support this approach, several empirical models of ultrashort photon pulse interactions with matter have been pursued previously \cite{num_md_i,md_haidu ,PhysRevE.69.051906,hau-riege:041902,plasma_mod}. However, owing to the large number of approximations employed, none have given a definitive answer to the basic question: is it possible to collect the scattering signal from an electronic ground state, to permit the reconstruction of the molecular structure? Within all these models, in fact, the primary interest is in the dynamics of atomic nuclei, while electrons are treated with very crude assumptions, ranging from simple rate equations to scattering cross sections, all of them derived in the framework of perturbative analysis of noninteracting electrons subjected to low external excitations. XFEL imaging, on the contrary, relies on the interplay between intense X-rays pulses and electron wavefunctions that, in turn, interact through nonlinear and extremely nonlocal (in space) functionals of the overall electron density \cite{PhysRevLett.52.997}
. An additional challenge to existing approaches also comes from the latest results obtained at LCLS, where ultrashort pulses less than $10$ fs length have been generated. Simulation indicates these pulses may be as short as $2$ fs with peak power levels of perhaps 500 GW. \cite{lclsfrish}.  The simulation provides novel insights in the internal structure of XFEL radiation, unveiling the presence of trains of incoherent energy bursts with mean time duration as short as $200$ as, whose interaction with matter cannot be modeled without a first-principle theory.\\
The frontier of atomic resolution imaging is still largely debated with the most problematic issues unsolved. Open questions include the photoionization and subsequent Coulomb explosion dynamics of a single molecule subjected to intense XFEL radiation (and in particular, if it could survive long enough to allow the observation of a useful diffraction signal) \cite{rev_xfel_chap}, what is the sample far field scattering pattern and if it encodes interference fringes with sufficiently high contrast to infer structural information on the material and, last but not least, what is the role of nonlinearity that, on the basis of recent theoretical calculations made on plasma models, is expected to affect the dynamics of high energy XFEL pulses \cite{fratalocchi:245132}.\\
The aim of this work is to answer to these questions. On one hand, to define the basis for present and future research in XFEL science, we develop an \emph{ab-initio} model describing the nonlinear interaction of intense XFEL beams with matter; on the other, to address the above-mentioned open questions we employ a numerical parallel approach with the highly scalable code GZilla (especially written and optimized for this problem) performing simulations with a variety of atoms and simple molecules. Our results are twofold:\\
i) we addressed the problem of sample radiation damage and demonstrate that the full photoionization of simple molecules occurs in a few ($<5$) of femtoseconds, with valence electrons escaping in $200-300$ attoseconds; by a further comparison with the photoionization time of the hydrogen atom, we elucidate the fundamental role of nonlinear electron interactions in the photoionization dynamics.\\
ii) we investigated the coherent imaging capabilities of XFEL sources, collecting snapshots of integrated far field diffraction patterns (as it would be retrieved by a standard camera), ions positions, electrons and electromagnetic energy density of molecules illuminated by ultrashort pulses (both in the femtosecond and attosecond regimes). We highlighted the existence of competitive dynamics on the femtosecond time scale, sustained by the strongly nonlinear nature of the electrons interactions, whose effect is to appreciably alter the electron density of the molecular system.\\
The main conclusion of this work is that, albeit some structural information information is still retrievable in the femtosecond domain, accurate single molecule imaging will be only possible in the sub-femtosecond regime.

\paragraph*{A first principle model and a state-of-the-art parallel solver}
To derive a first principle model of an ensemble of nonrelativistic atoms under the presence of a time-dependent electromagnetic field, we resort to the following quantum Hamiltonian $\mathbf{H}$ \cite{LoudonBook}:
\begin{equation}
  \label{eq:ham}
  \mathbf{H}=\int_V \frac{dV}{2}\bigg[\frac{\mathbf{n}_n(\mathbf{p}_n-Ze\mathbf{A})^2}{m_n}+\frac{\mathbf{n}_e(\mathbf{p}_e+e\mathbf{A})^2}{m}+\epsilon_0\mathbf{E}^2+\mu_0\mathbf{H}^2\bigg],
\end{equation}
with atomic nuclei (electrons) defined by the charge density operator $\mathbf{n}_n$ ($\mathbf{n}_e$), the momentum $\mathbf{p}_n$ ($\mathbf{p}_e$), the atomic number $Z$ (elementary charge $e$) and mass $m_n$ ($m$), while the electromagnetic (e.m.) field by the electric $\mathbf{E}$ and magnetic $\mathbf{H}$ field (or equivalently by the vector potential $\mathbf{A}$), with $\epsilon_0$ and $\mu_0$ being dielectric and magnetic constants, respectively. Equations of motion for both matter and e.m. field can be found directly from (\ref{eq:ham}) through the application of the Poisson bracket operator (supporting online material). At variance with previous theoretical and numerical studies, here all the quantities appearing in the quantum Hamiltonian (i.e., nuclei, electrons and e.m. field) are treated as dynamical variables, evolving through the nonlinear set of equations arising from the Hamiltonian (\ref{eq:ham}) itself (supporting online text).\\ 
 The numerical simulation of the dynamics arising from (\ref{eq:ham}) is accomplished with the parallel code GZilla (supporting text online), which combines classical Density Functional Theory (DFT) \cite{ES} for the computation of the system ground state, and TDDFT coupled to Finite-Difference Time-Domain (FDTD) \cite{TafloveBook} and Molecular Dynamics (MD) codes for the time-dependent analysis of (\ref{eq:ham}). GZilla has been parallelized with the Message-Passing-Interface (MPI) standard, which allows top performance in processor scalability, tested and verified on up to 4096 processors to date. The results presented have been derived by 15 million of single CPU computational hours, representing one of the longest simulations ever reported; thanks to the code scalability and to the availability of parallel computing resources, more than 1700 years of single cpu computations have been completed in less than 3 months.

\begin{figure}
\centering
\includegraphics[width=12cm]{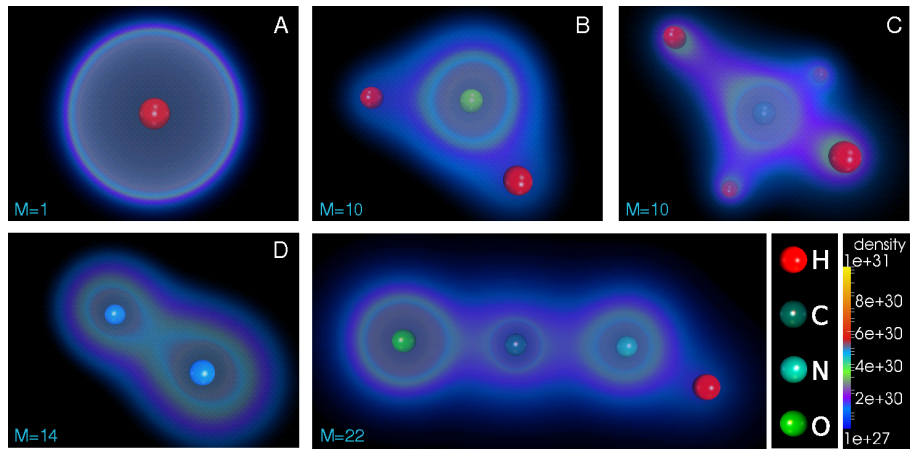}
\caption{
\label{fig1}
Three dimensional images of ground state molecules, with nuclei displayed as rigid spheres of arbitrary size, of (A) $H$, (B) $H_2O$, (C) $CH_4$, (D) $N_2$ and (E) $HNCO$ (the bottom-left part of each panel shows the corresponding number of electrons).
}
\end{figure}

\paragraph*{Setting the ground state}
In order to provide an adequate and experimentally interesting selection of molecules, we look at the four commonly-found elements of organic chemistry $(H,O,N,C)$ and study their stable configurations, ranging from the most elementary compound to the simplest combination involving all of them (Fig. \ref{fig1}). Ordered by an increasing degree of electronic complexity, we considered the hydrogen atom $H$ (Fig. \ref{fig1}-A), water $H_2O$ (Fig. \ref{fig1}-B), methane $CH_4$ (Fig. \ref{fig1}-C), nitrogen $N_2$ (Fig. \ref{fig1}-D) and finally the isocyanic acid $HNCO$ (Fig. \ref{fig1}-E) \cite{figs}. The chosen molecules exhibit a mean bond length of about $0.1$ nm, which matches well with the XFEL shortest wavelength ($\approx 0.1$ nm), and different sizes of external electron clouds, allowing the investigation of the interplay between nonlinearity and sample geometry at various spatial scales.

\begin{figure}
\centering
\includegraphics[width=5.5cm]{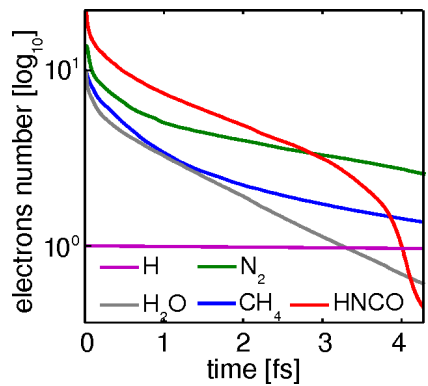}
\caption{
\label{fig2}
(A) Number of 'bounded' electrons versus time for $H$, $H_2O$, $CH_4$, $N_2$, and $HNCO$ molecules illuminated by a $100$ nm-waist Gaussian beam of cw XFEL radiation with power $P=150$ GW and wavelength $\lambda=0.4$ nm.
}
\end{figure}

\paragraph*{Molecules in high intensity e.m. field: radiation damage and photoemission rates}
The first series of \emph{ab-initio} simulations are devoted to the study  radiation damage of samples subjected to intense XFEL radiation, which is a key issue to calibrate XFEL experiments. The spatial distribution of the input source has been modeled with a Gaussian spatial profile, bounded to $100$ nm, a focusing which appears feasible with the existing technology. To investigate the interplay between nonlinearity and geometry, we first considered a continuous (cw) excitation. Parameters for wavelength and power have been chosen to match the XFEL in the high energy regime; in particular, for wavelength $\lambda$ well below $1$ nm, we considered the SASE 3 configuration of the European XFEL ($\lambda=0.4$ nm, power $P=150$ GW) \cite{xfele}. Figure (\ref{fig1}) shows the photoionization of $H$, $H_2O$, $CH_4$, $N_2$ and $HNCO$,  displaying the time evolution of the total number of electrons in the computational box (a cube of $1.5$ nm side for $N_2$ and $1$ nm for all the others). As seen in Fig. \ref{fig1}, the sample radiation damage is extremely high: with the only exception the $H$ atom, large ionization occurs in a few of femtoseconds with the $HNCO$ molecule exhibiting a radiation damage (percentage of electrons lost) of $98\%$ at a time $t=4$ fs, showing two different characteristic exponential ionization rates versus time. As a result, at the single femtosecond scale (much before a measurable Coulomb explosion of the nuclei), the dynamics is completely dominated by electrons shaking up; since the latter are the only source of scattered photons (nuclei, in fact, are too heavy to produce any appreciable scattering), any imaging-oriented XFEL application should be realized within ultrashort ($\le 1$ fs) pulses.\\
A second key observation comes from the different behavior of $H$, which indicates that nonlinearity, rather than geometry, outweigh the photoionization process. In fact, as displayed in Fig. \ref{fig2}A, the internal electrons of $H_2O$ and $HNCO$, whose wavefunctions have spatial extent comparable to those of $H$, escaped much before those of $H$ showing that electron nonlinear interactions actually dominate the system evolution. A physical interpretation comes from the theory of solitons \cite{KivsharBook}; the electron ground-state, in fact, being a solution of the dynamical equations arising from (\ref{eq:ham}) with constant amplitude and linear phase evolution, can be regarded as a a set of $M$ solitary waves interacting through a functional of the electron density (i.e., the $V_{xc}$ potential), which plays the role of a nonlinear Kerr-like response. Such nonlinearity is well known to foster processes with a considerable amount of radiation emission, as soon as solitary waves are perturbed from their ground state and forced to overlap. Therefore, when the XFEL pulse actually breaks in and upsets the atomic system, it initiates a process of radiation emission that gets amplified by the system nonlinear response through the interaction among electronic solitary waves. It is clear that this mechanism does not play a role for hydrogen, which has just one electron, resulting in a much lower photoionization rate with respect to heavier atoms.

\begin{figure}
\centering
\includegraphics[width=12cm]{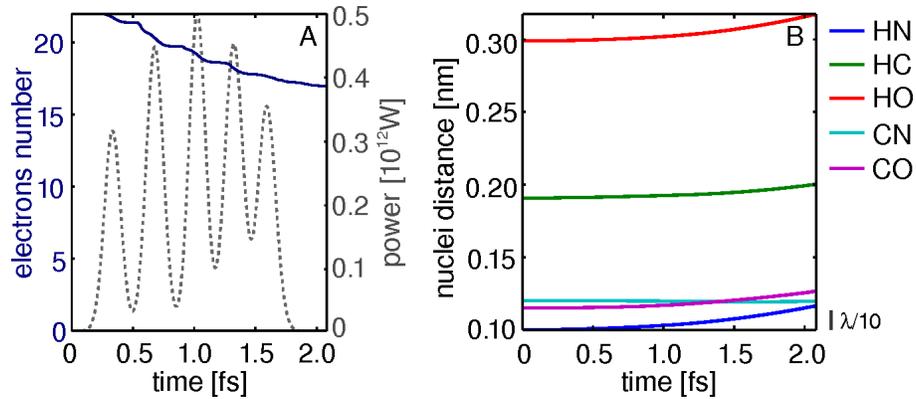}
\caption{
\label{fig3}
Time evolution of electrons number (A, solid lines) and nuclei distance (B) of $HNCO$ irradiated by a random train of incoherent energy bursts of peak power $P=500$ GW, wavelength $\lambda=0.15$ nm and overall time length T=$2$ fs (A, dashed line).
}
\end{figure}

\begin{figure}
\centering
\includegraphics[width=12cm]{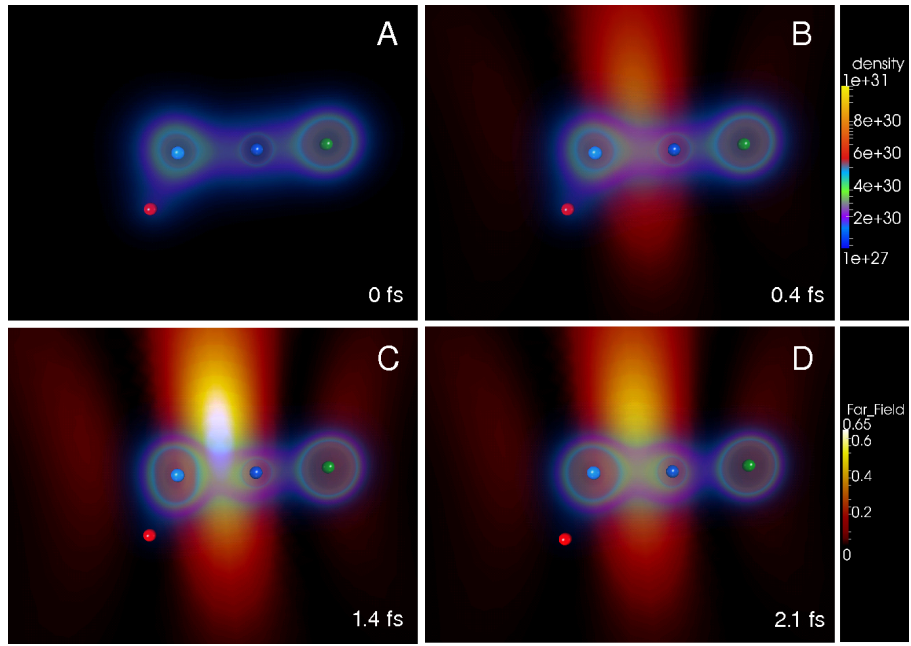}
\caption{
\label{fig4}
(A)-(D) Far-field scattered angular pattern (red to yellow colormap), nuclei position and electron density (blu to yellow colormap) time evolution of an $HNCO$ molecule irradiated by a short XFEL pulse of $2$ fs length, wavelength $\lambda=0.15$ nm and peak power $P=500$ GW (Fig. \ref{fig3} A dashed line). 
}
\end{figure}

\begin{figure}
\centering
\includegraphics[width=12cm]{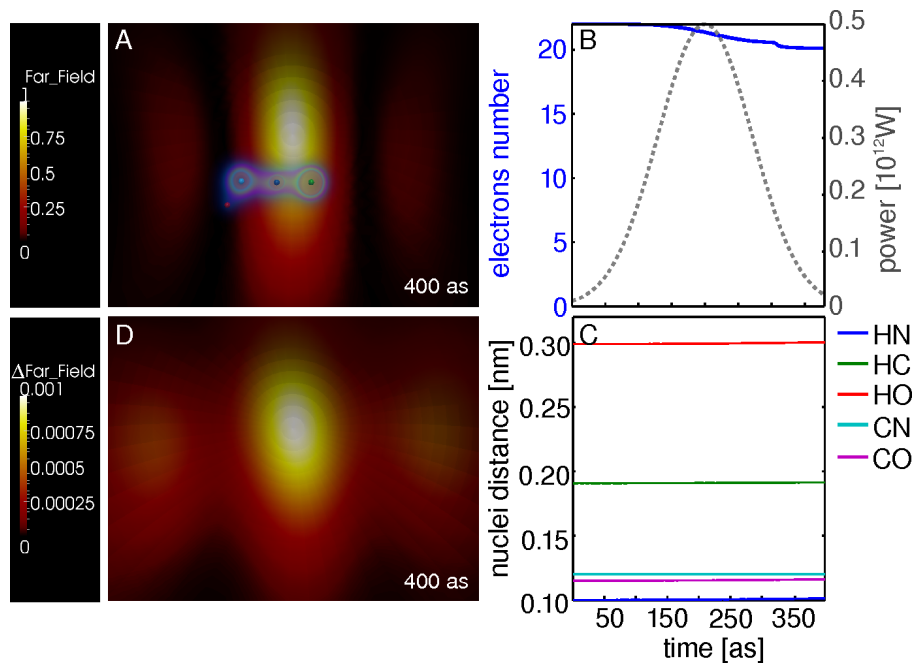}
\caption{
\label{fig5}
Attosecond diffraction results: (A) Angular Far-field, nuclei position/electron density, (B) photoionization and (C) nuclei distance evolution of an $HNCO$ molecule irradiated by a $400$ as XFEL pulse of wavelength $\lambda=0.15$ nm and peak power $P=500$ GW (B, dashed line); (D) Image difference between the far field of (A) and the one generated by the ground state of $HNCO$ in the absence of any radiation damage.
}
\end{figure}

\paragraph*{Single molecule imaging: a benchmark with the isocyanic acid HNCO}
The most challenging (and unsolved) issue concerns the possibility of recording the far field molecular interference image on a ÔcameraÕ that, can not employ any time gating and will detect scattered photons within the duration of the X-ray pulse. To answer to this question, we performed a series of simulations collecting the time domain far-field pattern \cite{TafloveBook}, integrated in time as it would be done by a standard CCD, scattered by an $HNCO$ molecule irradiated with an high energy XFEL pulse of Gaussian spatial distribution ($100$ nm spot-size) under different ultrashort regimes. We begin by considering the XFEL short pulse operation, recently characterized at the LCLS facility, which is expected to produce $2$ fs pulses with peak powers of $P=0.5$ TW at $\lambda=1.5$ nm \cite{lclsfrish}. This regime allows a direct comparison with the XFEL technology accessible in the near future. The set of Figs. (\ref{fig3})-(\ref{fig4}) provides a comprehensive analysis by showing the photoionization (Fig. \ref{fig3} A), the distance between the nuclei (Fig. \ref{fig3} B), the electron/nuclei dynamics and the angular far field (Movie S1 and movie snapshots Fig. \ref{fig4}) time evolution of an $HNCO$ molecule irradiated by a XFEL short pulse, constituted by a random series of Gaussian-shaped incoherent energy bursts (single burst length $\approx 200$ as, overall length $2$ fs), with peak power $P=500$ GW and wavelength $\lambda=0.15$ nm (Fig. \ref{fig3} A dashed line). As seen in Fig. \ref{fig3} A, each XFEL energy burst pushes the HNCO ionization a step forward, appreciably damaging the molecule much before the arrival of the main pulse peak (Fig. \ref{fig3} A, dashed line). At the point of maximum XFEL intensity ($\approx 1$ fs), in fact, the $HNCO$ molecule has lost three electrons; after the pulse peak, HNCO ionization proceeds at a slower rate and at the end the radiation damage reaches a value of $23\%$, with more than five electrons escaped from the molecule. Figure \ref{fig3} B illustrates the consequences of molecular damage on nuclei dynamics. In particular, the hydrogen atom is no longer bound to the structure (after $400$ as) and initiates the process of Coulomb explosion. The nuclei pairs $C-N$ and $C-O$, conversely, provide opposite dynamics: while the latter exhibit a tendency towards explosion, the former are not affected by this process and evolve with a constant bond length within the XFEL time window. In the framework of the previous nonlinear wave interpretation of the dynamics, such a different evolution is the hallmark of the generation of a robust multi-soliton cluster, with a consequent spatial reshaping of the electron density due to the nonlinear mixing of $C$ and $N$ electron waves. Movie S1 and Figs. \ref{fig4} A-D provide a clear illustration of this nonlinear process. At the beginning (t=0 fs), the carbon atom shows no preferential coupling to either nitrogen or oxygen, as witnessed by the symmetric density distribution which equally affects $C$ and $N$ neighbors. However, as soon as the hydrogen atom looses its coupling to the structure (Fig. \ref{fig4} C), the situation dramatically changes: the carbon and nitrogen atoms interact and evolve towards a robust multi-soliton cluster, as demonstrated by the strongly asymmetric density distribution which survives at the end of the XFEL pulse (Fig. \ref{fig4} D).\\
In conclusion, even by the use of short pulses and in the presence of negligible Coulomb explosion, the electronic radiation damage of molecular samples is still an issue: on one hand, it deprives the molecules of an appreciable number of electrons, on the other, it initiates a change of the electron density through the formation of preferred multi-soliton clusters, which tend to oppose sample ionization. The net effect of these two different dynamics (both sustained by the strongly nonlinear nature of the light-matter interaction process) is a qualitative and quantitative reshaping of the electron density which, in turn, leads to the inability to reconstruct the molecular structure from the far field diffraction pattern. In other words, although the few femtosecond time scale is short enough to avoid a significant Coulombian explosion of the nuclei, the sub-femtosecond ($\approx 0.5$ fs) electron density reshaping is still an issue for single molecule imaging experiments.\\ 
A general improvement is possible by moving to the attosecond domain, as expected on the basis of Fig. \ref{fig3} A, which predicts considerably smaller radiation damage after the first energy burst ($\approx 0.5$ fs). Figure \ref{fig5} summarizes the results of attosecond molecular diffraction: by employing a $400$ as long ultrashort XFEL pulse, the far field diffraction pattern (despite the short illumination) shows good fringe contrast (Fig. \ref{fig5} A), much superior to the pattern obtained in the fs regime (Fig. \ref{fig4} D), minimal radiation damage ($8.5\%$ as calculated from Fig. \ref{fig5}B) and no tendency toward nuclei explosion (Fig. \ref{fig5} C). Figure \ref{fig5}D, showing the difference between the previous far field diffraction pattern and the $HNCO$ diffraction from the ground state (i.e., in the absence of any molecular damage), highlights the minimal effects of radiation damage in the attosecond diffraction regime (Fig. \ref{fig5} A,D). The sub-femtosecond domain, in other words, is sufficiently fast to minimize the nonlinear interaction dynamics of electron matter waves (Fig. \ref{fig5}A) and to maintain all the nuclei bonded together (Fig. \ref{fig5} C). A computer reconstructed image from an attosecond diffraction pattern would contain all the information of the original electron density and will therefore be a true representation of the original sample.\\ 
In summary, in both short and ultrashort regimes the Coulomb explosion of the nuclei is not an issue for single molecule imaging, but the radiation damage (due to the high energy carried by the XFEL pulse) do offer some limitations to Angstrom-scale imaging. In particular, competitive dynamics originated and sustained by nonlinear electron interactions lead to a distortion of the electron density distribution in the XFEL femtosecond regime. Albeit some structural information is still retrievable by employing femtosecond XFEL pulses, a computer reconstructed image will no longer be a correct reproduction of the original sample due to the unbinding of some molecular nuclei and the formation of preferred multi-soliton clusters. Such dynamics are avoided in the attosecond domain, which is sufficiently faster then the characteristic scale of such effects, preserving the original electron distribution and the resulting diffraction pattern from loosing any important information.

\paragraph*{Dream or reality?}
With the opening of LCLS and the collection of its initial experimental results, the XFEL has become a reality of unprecedented importance. In this paper, we have derived an original and accurate description of XFEL beam interaction with matter, able to provide \emph{ab-initio} results and expected to settle the basis for present and future theoretical investigations in XFEL science. When applied to the one of the most promising applications of XFEL sources, our results predict that some work still needs to be done in order to reach the goal of single molecule imaging and Angstrom scale microscopy, but we foresee that such frontiers may be just behind corner, especially in conjunction with the fast theoretical and experimental advances in the field of attosecond physics \cite{krausz:163}.



\begin{scilastnote}
\item This research is done in the framework of \emph{The SolarPaint project} (www.solarpaintproject.org), an activity funded by  Award No. KUK-F1-024-21 (2009/2012) made by King Abdullah University of Science and Technology (KAUST). We are indebted to KAUST (Aron Ahmadia,  Moody Altamimi, Bruce Guile, Dinesh Kaushik, David Keyes, Laura Meany, Richard Orme) and KAUST-IBM partnership (Chris Beyer, Pavan Kumar, Cindy Mestad, Fred Mintzer, Dave Singer, Julie Sherlock) for the large parallel resources provided through KAUST Shaheen, IBM WatsonShaheen and IBM Rochester Blue-Gene supercomputers. Code development has been done at the Max-Plank IBM Power6 system, thanks to the Distributed European Infrastructure for Supercomputing Applications (DEISA) European award given to the Ab-initio Coulomb Explosion Simulations at X-rays (ACES-X) project. We are very grateful to the CINECA team (Cristiano Calonaci, Andrew Emerson, Giovanni Erbacci) for the parallel support in the enabling stage of ACES-X. We finally acknowledge illuminating discussions with Massimo Altarelli (DESY European XFEL) and Jerry Hastings (Stanford LCLS) concerning the practical capabilities of XFEL light sources.
\end{scilastnote}

\textbf{Supporting Material}\\
SOM text\\
Movie S1\\
References S2, S3, S4


\clearpage

\clearpage

\section*{Supporting Material}

\section{Supporting Text}
\label{sec:txt}

\subsection{Equations of motion}
Atomic nuclei, due to their extremely small size ($\approx 10^{-15}$m) with respect to hard X-ray wavelengths $(>10^{-10}$m), are seen as point objects from the XFEL, hence, the quantum expectation of the nuclei density $\langle \Psi |\mathbf{n}_n(\mathbf{r})|\Psi\rangle=\sum_i\delta(\mathbf{r}-\mathbf{R}_i)$ (averaged over the many body state $|\Psi\rangle=|\psi\rangle|\chi\rangle$ composed by electrons $|\psi\rangle$ and nuclei $|\chi\rangle$) factorizes into a series of Dirac delta centered at the nuclei positions $\mathbf{R}_i$.  The latter, and their conjugate momenta $\mathbf{P}_i$, evolve according to the following Poisson brackets:
\begin{eqnarray}
  &&\frac{\partial \mathbf{R}_i}{\partial t}=\{\mathbf{R}_i,\langle \Psi | \mathbf{H}|\Psi\rangle\},\nonumber\\
  &&\frac{\partial \mathbf{P}_i}{\partial t}=\{\mathbf{P}_i,\langle \Psi | \mathbf{H}|\Psi\rangle\},
\end{eqnarray}
with $i=1,...,N$. After some algebra, we end with \footnote{Throughout this article, we assume to work in the Coulomb gauge.}:
\begin{equation}
  \label{eq:io}
   M_i\frac{\partial ^2\mathbf{R}_i}{\partial t^2}=Z_ie\big[(\mathbf{E}_{Ti}+\mathbf{v}_i\times\mathbf{B}_i)-\nabla_{\mathbf{R}_i}\phi(\mathbf{R}_i)\big],
\end{equation}
with $\mathbf{E}_{Ti}=\mathbf{E}_T(\mathbf{R}_i)$ [the subscript $_T$ denotes the transverse part] and the potential $\phi$:
\begin{equation}
  \label{eq:io1}
  \phi(\mathbf{R}_i)=\frac{e}{4\pi\epsilon_0}\bigg(\sum_{j\neq i}\frac{Z_j}{|\mathbf{R}_i-\mathbf{R}_j|}-\int\mathrm{d\mathbf{r}}\frac{n(\mathbf{r})}{|\mathbf{R}_i-\mathbf{r}|}\bigg),
\end{equation}
being $Z_i$ the atomic number of the $i-$th nucleus and $n$ the quantum expectation of the electron density.\\
The evolution of the electron many body state $|\psi\rangle$, in the Schr\"odinger representation, is described by:
\begin{equation}
  i\hbar\frac{\partial}{\partial t}|\psi\rangle=\mathbf{H}|\psi\rangle
\end{equation}
The latter, in its direct form, is unsuitable for a numerical integration due to a dimensionality bottleneck problem. The memory workload for storing the wavefunction $\psi$, in fact, grows exponentially as $L^M$ for an $M-$body system with size $L$. By using a tiny grid of $128$ points in each spatial dimension (typical simulations fall in the range $256-1024$ points), more than $60$ millions of terabytes would be required for storing the wavefunction of just $3$ electrons, which makes the problem unmanageable even in this simple situation. A more convenient formulation is provided by Time Dependent Density Functional Theory (TDDFT) \cite{TDDFT,ES}, which rewrites the single many-body problem (\ref{eq:el}) into a series of one-body problems coupled via a nonlinear potential $V_{xc}$ (the \emph{exchange-correlation} potential), function of the electron density $n$, which describes the overall interaction among electrons. This formulation scales as $L\cdot M$, and can be handled in a sufficiently-large supercomputing environment. As in Hartree-Fock theory and noninteracting quantum many-body dynamics \cite{ES,advqm}, TDDFT writes down the many body state as a Slater determinant:
\begin{equation}
  |\psi(\mathbf{r_1},...,\mathbf{r}_M)\rangle=\frac{\det({S})}{M!}
\end{equation}
with $S_{ij}=|\psi_j(\mathbf{r}_i)\rangle$, and $\psi_j=\langle \mathbf{r}|\psi_j\rangle$ playing the role of a single orbital whose evolution, according to Eq. (\ref{eq:ham}), is found to be:
\begin{equation}
  \label{eq:el}
  i\hbar\frac{\partial\psi_j}{\partial t}=\bigg[-\frac{\hbar^2}{2m}\nabla^2+\frac{e^2}{4\pi\epsilon_0}\bigg(\int d\mathbf{r}'\frac{n(\mathbf{r}')}{|\mathbf{r}-\mathbf{r}'|}-\sum_i\frac{Z_i}{|\mathbf{R}_i-\mathbf{r}|}\bigg)+V_{xc}+\frac{e^2}{2m}\mathbf{A}^2-i\hbar\frac{e}{m}\mathbf{A}\cdot\nabla\bigg]\psi_j,
\end{equation}
being $\mathbf{A}$ the electromagnetic vector potential and $j$ running from $1$ to $M$.\\
The equations of motion of the electromagnetic fields, finally, are obtained from:
\begin{eqnarray}
  &&\frac{\partial \mathbf{q}}{\partial t}=\{\mathbf{q},\langle \Psi | \mathbf{H}|\Psi\rangle\},\nonumber\\
  &&\frac{\partial \mathbf{p}}{\partial t}=\{\mathbf{p},\langle \Psi | \mathbf{H}|\Psi\rangle\},
\end{eqnarray}
with $\mathbf{p}=-\epsilon_0\mathbf{E}$ and $\mathbf{q}=\mathbf{A}$ acting as momentum and position, respectively. Equations (\ref{eq:em}) yield the following nonlinear Maxwell's set:
\begin{eqnarray}
  \label{eq:em}
&&\frac{\partial\mathbf{E}}{\partial t}=\frac{1}{\epsilon_0}\bigg[\nabla\times\mathbf{H}+\mathbf{J}\bigg],\nonumber\\
&&\frac{\partial\mathbf{H}}{\partial t}=-\frac{\nabla\times\mathbf{E}}{\mu_0},\nonumber\\
&&\frac{\partial \mathbf{A}}{\partial t}=-\mathbf{E}_\mathrm{T},\nonumber\\
&&\nabla\cdot\mathbf{E}_T=\nabla\cdot\mathbf{B}=\nabla\cdot\mathbf{A}=0.
\end{eqnarray}
 with a source $\mathbf{J}$, representing the classical current of electrons and nuclei:
\begin{equation}
  \label{eq:j}
  \mathbf{J}=e\bigg[\sum_{i=1}^N\frac{Z_i\mathbf{P}_i}{m_i}\delta(\mathbf{r}-\mathbf{R}_i)-\frac{1}{m}\langle\psi|\mathbf{n}_e\mathbf{p}_e|\psi\rangle-\frac{e\mathbf{A}}{m}n\bigg],
\end{equation}
The set of equations represented by (\ref{eq:io})-(\ref{eq:io1}), (\ref{eq:el}) and (\ref{eq:em})-(\ref{eq:j}) forms the complete theoretical model of XFEL interaction with matter. The first principle XFEL model integrates molecular dynamics [Eqs. (\ref{eq:io})-(\ref{eq:io1})], Maxwell's [Eqs. (\ref{eq:em})-(\ref{eq:j})] and hydrodinamics-like [Eqs. (\ref{eq:el})] equations in nonlinearly coupled platforms, which evolve on comparable time scales. A further complication in the simulation of the set (\ref{eq:io})-(\ref{eq:io1}), (\ref{eq:el}) and (\ref{eq:em})-(\ref{eq:j}) originates from the mixing of both transverse and global quantities in Maxwell's equations (\ref{eq:em})-(\ref{eq:j}), as well as the extremely small contribution arising from the scattered current $\mathbf{J}$, which generate radiation fields several orders of magnitude lower than the XFEL beam. More specifically, according to the Helmholtz theorem (S2), the calculation of the transverse part of a generic vector field requires the evaluation of a 3D vector integral (which is very time consuming), while the extrapolation of such a small scattered field poses challenges even to the advanced total field/scattered field (TFSF) formulation for solving Maxwell's equations (able to isolate the scattered field contribution stemming from a single numerical grid). We addressed these problems by defining two different sets of Maxwell's equations, one dealing with the transverse field and one with the scattered contributions, exploiting the division naturally arising from the use of the Coulomb gauge (S2). In particular, we considered the following equations modeling the evolution of the transverse beam:
\begin{eqnarray}
  \label{eq:em1}
&&\frac{\partial\mathbf{E}_T}{\partial t}=\frac{1}{\epsilon_0}\nabla\times\mathbf{H}_T,\nonumber\\
&&\frac{\partial\mathbf{H}_T}{\partial t}=-\frac{\nabla\times\mathbf{E}_T}{\mu_0},\nonumber\\
&&\frac{\partial \mathbf{A}}{\partial t}=-\mathbf{E}_\mathrm{T},\nonumber\\
&&\nabla\cdot\mathbf{E}_T=\nabla\cdot\mathbf{B}_T=\nabla\cdot\mathbf{A}=0.
\end{eqnarray}
and the next set for the dynamics of the scattered field:
\begin{eqnarray}
  \label{eq:em2}
&&\frac{\partial\mathbf{E}}{\partial t}=\frac{1}{\epsilon_0}\bigg[\nabla\times\mathbf{H}+\mathbf{J}\bigg],\nonumber\\
&&\frac{\partial\mathbf{H}}{\partial t}=-\frac{\nabla\times\mathbf{E}}{\mu_0},
\end{eqnarray}
which is triggered by the current $\mathbf{J}$ generated by electrons and nuclei through (\ref{eq:j}) and depending on the vector potential $\mathbf{A}$ of (\ref{eq:em1}). Such a transverse field/scattered field formulation is particularly convenient since it allows to safely neglect the transverse scattered contributions in (\ref{eq:em1}), as they are orders of magnitude lower than the ultraintense XFEL beam, yielding a set of equations that can be integrated very efficiently (no mixing with transverse and total fields) and which provide an accurate description of the scattered field arising from the molecules through (\ref{eq:em2}).

\subsection{GZilla code implementation details}
GZilla features one of the most comprehensive simulator ever realized, composed to date of more than 120000 lines of parallel code. Its analysis capabilities range from DFT ground state, to time-dependent quantum molecular dynamics in the presence of electromagnetic fields of arbitrary form. Time-dependent far-field pattern analysis is also provided. GZilla implementation has been developed by taking into account state-of-the-art methods and algorithms. In particular:\\
Maxwell's equations (\ref{eq:em})-(\ref{eq:j}) are solved by the FDTD technique within a Cartesian Yee grid with time marching Leapfrogging, Uniaxial Perfectly Matched Layers (UPML) and Total Field Scattered Field (TFSF) source formulation. The far-field pattern is calculated by the time-domain technique \cite{TafloveBook}. Code parallelization is provided through spatial and time domain decompositions.\\
The time evolution of Schr\"odinger equations (\ref{eq:el}) is conversely committed to an original, unconditionally stable Crank Nicholson propagator designed with graph theoretical approaches to minimize communications among processors and an original theory of Perfectly Matched Layers (PML) to absorb outgoing electron waves, combined with a Galerkin-based geometric Multigrid (S3) for the evaluation of the density integral appearing on the r.h.s of Eq. (\ref{eq:el}). Code parallelization is provided through a spatial domain decomposition strategy.\\
Equations (\ref{eq:io})-(\ref{eq:io1}), finally, are solved by a velocity-verlet time stepping algorithm \cite{ES}, parallelized by an original communicator splitting technique.\\
The ground state DFT analysis is derived through a Rayleigh Quotient Multigrid (RQMG) eigenvalue solver (S4), within a Self-Consistent Field (SCF) iteration procedure \cite{ES} and an original guaranteed-reduction charge-mixing scheme.

\end{document}